\documentclass{aa}
\usepackage{graphicx}
\usepackage{txfonts}
\begin{document}
\title{SCUBA sub-millimeter observations of gamma-ray bursters}

\subtitle{III. GRB 030329: The brightest sub-millimeter afterglow to date}

\author{I.A. Smith\inst{1}
\and R.P.J. Tilanus\inst{2}
\and N. Tanvir\inst{3}
\and R.A.M.J. Wijers\inst{4}
\and P.~Vreeswijk\inst{5}
\and E. Rol\inst{6}
\and C. Kouveliotou\inst{7,8}}

\offprints{I. Smith}

\institute{
Department of Physics and Astronomy, Rice University, 
6100 South Main, MS-108, Houston, TX 77005-1892 USA\\
\email{iansmith@rice.edu}
\and Joint Astronomy Centre, 660 N. Aohoku Place, Hilo, HI 96720 USA
\and Centre for Astrophysics Research, University of Hertfordshire, 
College Lane, Hatfield, Herts AL10 9AB, UK
\and Astronomical Institute `Anton Pannekoek', 
University of Amsterdam and Center for High-Energy Astrophysics,\\
Kruislaan 403, 1098 SJ Amsterdam, The Netherlands
\and European Southern Observatory, Alonso de C\'ordova 3107, 
Casilla 19001, Santiago 19, Chile
\and Department of Physics and Astronomy, University of Leicester, 
Leicester LE1 7RH, UK
\and NASA Marshall Space Flight Center, SD-50, NSSTC, 
320 Sparkman Drive, Huntsville, AL 35805 USA
\and Universities Space Research Association}

\date{Received ; accepted }

\abstract{
We discuss our ongoing program of Target of Opportunity (ToO)
sub-millimeter observations of gamma-ray bursts (GRBs) 
using the Sub-millimetre Common-User Bolometer Array (SCUBA) 
on the James Clerk Maxwell Telescope (JCMT).
In this paper, we present all the ToO observations of GRB 030329.
This was by far the brightest sub-millimeter afterglow seen to date.
The flux density at 850~$\mu$m was approximately constant 
up to a break that took place $\sim 7$ days after the burst.
This was consistent with being a jet break.
The 850~$\mu$m results agree with those at longer wavelengths that 
show a brighter flux $\sim 7$ days after the burst, right at the 
time of the break.
No short-lived large-scale brightenings were detected in the sub-millimeter 
light curve.
However, the 850~$\mu$m light curve may have had a drop 
$\la 16$ days after the burst.
The peak of the afterglow emission was at $\sim 90$ GHz in the
days before the break in the light curve.
A simple modeling is consistent with the spectral indices remaining the
same as the afterglow evolved, with the breaks in the spectrum
moving to longer wavelengths at later times and the flux at the 
peak falling.
No significant sub-millimeter emission was detected from the 
host galaxy.
}

\maketitle

\section{Introduction}

The discovery of localized transients in the error boxes of
gamma-ray burst (GRB) sources has led to intense multi-wavelength
campaigns that have revolutionized our understanding of these sources.
For reviews see Van Paradijs et al. (\cite{scuba:vanp00}) and 
M\'esz\'aros (\cite{scuba:mes02}).

Sub-millimeter observations form a key element of
the multiwavelength observations of the afterglow.
They provide ``clean'' measures of the source intensity, unaffected 
by scintillation and extinction.
We have therefore been performing Target of Opportunity (ToO)
sub-millimeter observations of GRB counterparts using the 
Sub-millimetre Common-User Bolometer Array (SCUBA) 
on the James Clerk Maxwell Telescope (JCMT) on Mauna Kea, Hawaii.

Including the results shown here, SCUBA has so far performed ToO 
observations of 23 bursts (Smith et al. \cite{scuba:smith99}, 
\cite{scuba:smith00}, \cite{scuba:smith01}, \cite{scuba:smith05}; 
Bloom et al. \cite{scuba:bloom98}; 
Galama et al. \cite{scuba:gal99nat}; Kulkarni et al. \cite{scuba:kfs99};
Frail et al. \cite{scuba:frail00}, \cite{scuba:frail02}, \cite{scuba:frail03};
Berger et al. \cite{scuba:berger00}, \cite{scuba:berg03a};
Yost et al. \cite{scuba:yost02}).
A complete summary of the previous SCUBA results and the motivations for 
performing sub-millimeter observations of the afterglows and host 
galaxies is given in Smith et al. (\cite{scuba:smith05}).
In this paper, we present all the SCUBA ToO observations of GRB 030329.

\section{Multiwavelength observations of GRB 030329}

The long ($> 100$ sec) GRB 030329 was detected by {\it HETE-2} and
several spacecraft in the Interplanetary Network at 20030329.484 UT 
(Vanderspek et al. \cite{scuba:van04}).
The intense ionizing flux from the burst produced a sudden disturbance 
to the Earth's ionosphere (Schnoor et al. \cite{scuba:swf03}).

The redshift was determined to be 0.1685 (Greiner et al. \cite{scuba:gre03a}), 
corresponding to a luminosity distance of $\sim 810$ Mpc.
This was one of the closest GRBs localized to date.
Although its intrinsic luminosity was relatively low, the fact that
it was nearby made it one of the brightest bursts and afterglows ever 
recorded.

GRB 030329 was an X-ray rich burst, and a distinct, bright, soft X-ray 
component may have been present during the burst.
The X-ray afterglow seen by the {\it Rossi X-Ray Timing Explorer}
and {\it XMM-Newton} was bright 
(Marshall \& Swank \cite{scuba:ms03}; Marshall et al. \cite{scuba:mms03};
Tiengo et al. \cite{scuba:tiengo03}, \cite{scuba:tiengo04}).
However, the X-ray afterglow light curve was sparsely sampled.
It might extrapolate back to match the soft X-ray tail seen 
during the burst.
There appeared to be a break in the decay of the X-ray light curve 
$\sim 0.5$ days after the burst, and there may have been a flattening of 
the X-ray light curve $\sim 40$ days after the burst.

Optical observations simultaneous with the burst did not detect the 
source, placing a limit of $V \sim 5.5$ on the reverse shock emission
(Torii et al. \cite{scuba:torii03}).
However, a bright optical afterglow was found, with $R \sim 12.4$ at 
67 minutes after the burst (e.g. Price et al. \cite{scuba:price03};
Sato et al. \cite{scuba:sato03}; Uemura et al. \cite{scuba:uki03};
Urata et al. \cite{scuba:urata04}).

The optical afterglow was polarized at the $0.3 - 2.5$\% level, and the 
polarization was variable on time scales down to hours 
(Greiner et al. \cite{scuba:gre03b}).
The radio afterglow was not significantly polarized
(Finkelstein et al. \cite{scuba:fink04}; Taylor et al. \cite{scuba:tay05}).

The optical afterglow decay can broadly be described using a 
broken power law with a break $\sim 3 - 8$ days after the burst
(Lipkin et al. \cite{scuba:lip04}).
However, extensive observations of the optical afterglow light curve found a 
complex evolution of bumps and wiggles superimposed on the overall fading
(e.g. Greiner et al. \cite{scuba:gre03b}; Smith et al. \cite{scuba:dsmith03}; 
Matheson et al. \cite{scuba:mat03}; Bloom et al. \cite{scuba:bloom04}; 
Lipkin et al. \cite{scuba:lip04}).
These could have been due to structures in the forward and/or reverse
shocks, or repeated energy injection from the central engine
(e.g. Sari \& M\'esz\'aros \cite{scuba:sm00};
Granot et al. \cite{scuba:gnp03}),
or inhomogeneities in the ambient medium that the fireball 
expanded into (e.g. Berger et al. \cite{scuba:berger00}).
The interpretation of the various features in the light curve
is therefore complicated.

The radio and millimeter afterglows were the brightest recorded
for any burst to date (Berger et al. \cite{scuba:berg03b};
Sheth et al. \cite{scuba:sheth03}; Kuno et al. \cite{scuba:kuno04};
Taylor et al. \cite{scuba:tay04}; Kohno et al. \cite{scuba:kohno05}).
The radio and millimeter light curves were initially approximately flat 
or slowly rising.
After this, they broke to a power law $F_\nu \propto t^{-\delta}$ 
with a steep decay of $\delta \sim 2$.
This is consistent with a jet break 
(e.g. Sari et al. \cite{scuba:sari99}).
The light curves for the observations from 43 to 250 GHz all broke at 
$\sim 8$ days after the burst.
In the single broken power law model, this would suggest that the jet 
break took place at $\sim 8$ days after the burst rather than at 
$\sim 3$ days which is allowed from the optical observations.
The peak occurred at increasingly later times for longer wavelengths, 
for example, it was at $\sim 30$ days at 4.86~GHz.
These longer wavelengths were below the synchrotron self-absorption
frequency ($\nu_a$) and/or below the typical synchrotron 
frequency for the lowest energy electron in the power law ($\nu_m$).
Interstellar scintillation was an increasingly important factor 
below 15 GHz for the early observations when the source was compact.
At late times, the radio light curve showed an achromatic flattening
which is consistent with the blast wave becoming transrelativistic
(Frail et al. \cite{scuba:frail05}).

A more complex model for the multiwavelength emission used breaks in 
the early optical afterglow decay at $\sim 0.25$ and $\sim 0.5$ days
(e.g. Sato et al. \cite{scuba:sato03}; Burenin et al. \cite{scuba:bur03}).
The first break could be due to the cooling break frequency ($\nu_c$) passing 
through the optical band.
The second break would be similar to the one seen in X-rays;
since it is achromatic, this suggests it is a jet break.
However, an additional source of optical emission would then be
required $\sim 1$ day after the burst.
A possible model for the multiwavelength afterglow then uses two jets 
with different opening angles (Berger et al. \cite{scuba:berg03b};
Sheth et al. \cite{scuba:sheth03}; Lipkin et al. \cite{scuba:lip04};
Tiengo et al. \cite{scuba:tiengo04}).
The initial part of the optical and X-ray afterglow would come 
from a narrow ultra-relativistic jet.
The late X-ray afterglow and optical brightening would come from a wide, 
mildly relativistic jet, with a jet break $\sim 10$ days after the burst.
The plateau seen at longer wavelengths before the break could then
be a convolution of the falling flux density (as $t^{-1/3}$) from 
the narrow angle jet and a rising flux density (as $t^{1/2}$) from 
the wide angle jet.
Although this model describes the overall trends in the light curves, 
the multiple bumps and wiggles in the optical light curve still remain 
to be explained.

Beginning $\sim 6$ days after the burst it became evident that
there was an underlying component from the Type Ic supernova SN 2003dh.
The broad spectral features indicated a large expansion velocity 
similar to the Type Ic hypernovae SN 1998bw and SN 1997ef
(Stanek et al. \cite{scuba:stan03}; Hjorth et al. \cite{scuba:hjorth03};
Kawabata et al. \cite{scuba:kaw03}; Matheson et al. \cite{scuba:mat03};
Mazzali et al. \cite{scuba:maz03}; Kosugi et al. \cite{scuba:kos04}). 
This was direct spectroscopic evidence that at least some 
classical GRBs originate from core-collapse supernovae.
The SN 2003dh observations also indicated that the explosion was
aspherical.

The host is believed to be a low to moderate metallicity starburst 
dwarf galaxy with $V \sim 22.7$ and $R > 22.5$
(Fruchter et al. \cite{scuba:fruct03}; Matheson et al. \cite{scuba:mat03}).

\begin{table*}
\begin{minipage}[t]{\columnwidth}
\caption{SCUBA 850~$\mu$m (350 GHz) afterglow observations
of GRB 030329.}
\label{table:all850}      
\centering 
\renewcommand{\footnoterule}{}
\begin{tabular}{l l l l l l l l}
\hline\hline       
Burst       & \multicolumn{2}{c}{SCUBA observing times} & Time since   & Integration & $\tau_{850}$ & Afterglow 850 $\mu$m \\
            & Start                & Stop               & burst (days) & time (sec)  &              & flux density (mJy)   \\
\hline                    
GRB 030329  & 20030403.383 & 20030403.414 & \phantom{1}4.914 & 1800 & 0.623 & $\phantom{-}37.8 \pm 4.5$ \\
            & 20030404.369 & 20030404.401 & \phantom{1}5.901 & 1800 & 0.467 & $\phantom{-}27.6 \pm 2.9$ \\
            & 20030405.269 & 20030405.302 & \phantom{1}6.801 & 1800 & 0.356 & $\phantom{-}33.9 \pm 2.8$ \footnote{Not corrected for small pointing error} \\
            & 20030409.222 & 20030409.266 &           10.760 & 1836 & 0.433 & $\phantom{-}10.9 \pm 5.1$ \\
            & 20030415.223 & 20030415.254 &           16.754 & 1800 & 0.300 & \phantom{1}$-0.2 \pm 1.9$ \\
            & 20030416.339 & 20030416.370 &           17.870 & 1800 & 0.192 & \phantom{1}$-1.9 \pm 2.7$ \\
\hline                  
\end{tabular}
\end{minipage}
\end{table*}

\section{SCUBA observing details}

SCUBA is the sub-millimeter continuum instrument for the JCMT
(for a review see Holland et al. \cite{scuba:hol99}).
The observing, calibration, and reduction techniques used for
GRB 030329 were the same as has been used on other bursts and
are described in detail in Smith et al. 
(\cite{scuba:smith99,scuba:smith01,scuba:smith05}).

SCUBA uses two arrays of bolometers to simultaneously observe the same
region of sky.
The arrays are optimized for operations at 450 and 850 $\mu$m.
Photometry observations are made using a single pixel of the arrays.
The other bolometers in the arrays are used to perform a good sky noise 
subtraction (Archibald et al. \cite{scuba:arch02}).
During an observation, the secondary is chopped between the
source and sky.
The term ``integration time'' always refers to the ``on+off''
time, including the amount of time spent off-source.
An 18 sec integration thus amounts to a 9 sec on-source observation time.
A typical measurement consists of 50 integrations of 18 seconds;
we refer to this as a ``run.''
Each observation of a source in general consists of several such
runs, with focus, pointing, and calibration observations in between.

Version 1.6 of SURF\footnote{SURF is distributed by the Starlink Project.}
(Jenness \& Lightfoot \cite{scuba:jen00}) was used to combine the 
integrations, remove anomalous spikes, flatfield the array, and 
subtract the signal from the sky bolometers.
The zenith sky opacity was determined using ``skydips'' in which
the sky brightness temperature was measured at a range of elevations.
This was used to apply atmospheric extinction corrections to the 
observed target fluxes.
At least one standard calibration target was observed each night to 
determine the absolute flux of the GRB.

A typical integration time of 2 hours gives an rms $\sim 1.5$~mJy at 
850 $\mu$m.
However, the sensitivity depends significantly on the weather and
the elevation of the source; since our ToO observations are done
on short notice, sometimes these factors are less than ideal.
Based on observed variations of the gain factor and signal levels we 
estimate typical systematic uncertainties in the absolute flux calibrations 
of 10\% at 850 $\mu$m.
In general the rms errors of the observations presented here are larger than 
this uncertainty.

The pointing of the JCMT is checked several times during the night 
to ensure that it is reliable.
The pointing accuracy is usually a few arcsec.
However, an error in the track model used between 2000 August 25 and
2003 April 25 resulted in pointing errors that were non-negligible
(shifts larger than $4\arcsec$) for targets with elevations above 
$60\degr$ and over small ranges of azimuth.
Since the 850 $\mu$m bolometric pixel has a diffraction limited resolution
of $14\arcsec$, the target remains well within the bolometric pixel.
However, there will be an error in the flux that is measured.
For GRB 030329, the only problem was with a pointing calibration 
observation of 3C 273 on 2003 April 5.
However, this did not significantly affect the result for that day.

\section{SCUBA observations of GRB 030329}

Unfortunately, SCUBA was not on the telescope at the time of the burst
and it was several days before the first SCUBA observation could be made.
The weather was not good for most of the observations leading to larger
rms uncertainties than normal.
Despite these difficulties, GRB 030329 was still easily detected 
at 850~$\mu$m in the initial observations.

Table~\ref{table:all850} summarizes all the 850 $\mu$m SCUBA 
ToO observations for GRB 030329.
These expand on and supercede the results in Hoge et al. 
(\cite{scuba:hoge03}).
The start times are the times when the GRB was first observed on
each day, and the stop times are when the last observation of the 
GRB was completed.
The time since the burst uses the mid-point between these start and
stop times (although it should be noted that focus, pointing, and 
calibration observations may have also taken place in between the 
start and stop times); the time is the elapsed time in the Earth 
frame, not the rest frame of the host galaxy.
The ``integration time'' for each observation is the ``on+off'' time; 
only half of this is spent on-source.
The zenith optical depth at 850~$\mu$m ($\tau_{850}$) is given for
the time of the observation.

The $\sim 30$ mJy source initially detected for GRB 030329 was by 
far the brightest SCUBA afterglow seen to date. 
For all the previous ToO observations performed by SCUBA, 
the 850 $\mu$m flux density from the afterglow and/or reverse 
shock was $< 10$ mJy.

\subsection{SCUBA 850~$\mu$m light curve}

\begin{figure}
\centering
\includegraphics[bb=60 150 570 660,width=8.5cm,clip]{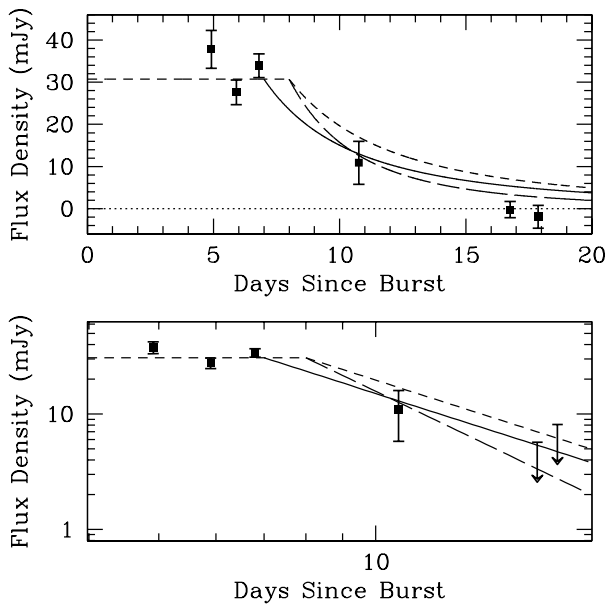}
\caption{The 850~$\mu$m light curve for GRB 030329.
The same data and models are plotted in the linear-linear (top)
and log-log (bottom) versions, except that the last two data
points are plotted as $3 \sigma$ upper limits in the latter.
The initial dashed curve is a constant with $F_\nu = 30.7$~mJy.
This breaks to $F_\nu \propto t^{-\delta}$, with the break
at 8 days and $\delta = 2$ (short dashed), the break at 8 days 
and $\delta = 3$ (long dashed), and the break at 7 days and
$\delta = 2$ (solid).}
\label{figure1}
\end{figure}

The 850~$\mu$m light curve is shown in Figure~\ref{figure1}.
This is similar to the observations at longer wavelengths.
The flux density is approximately constant initially, before
breaking with a rapid decay.
There is no evidence for an underlying dusty host galaxy.

The weighted mean for the first three SCUBA observations was 
$30.7 \pm 1.8$ mJy at 850~$\mu$m.
This gives the constant dashed curve shown in Figure~\ref{figure1} 
for the first week.

Unfortunately, we do not have high quality observations during the 
decay phase, due to the terrible weather at that time.
However, the lack of detection 17 days after the burst 
allows us to use simple qualitative models to constrain the break 
time and/or the decay slope.

The short dashed curve in Figure~\ref{figure1} shows a break at 8 days 
after the burst and a decay of $\delta = 2$.
This would predict a flux density of 7.0 mJy at 16.8 days after the
burst.
This is therefore inconsistent with our non-detection at the 
$3.7 \sigma$ level.

To maintain $\delta = 2$ for the whole decay, it would be necessary 
for the break to occur earlier.
The solid curve in Figure~\ref{figure1} places the break at 7 days.
This is just consistent with our $3 \sigma$ limit at 16.8 days.

To have the jet break occur at later times, it would be necessary to have 
a faster decay.
For example, the long dashed curve in Figure~\ref{figure1} uses 
a break at 8 days and $\delta = 3$.
Such a steep decay (on average) has not been seen in the other longer 
wavelength observations of this burst.
However, as shown in \S 4.2, it is possible that a break
moved through the sub-millimeter band between 7 and 17 days after
the burst, and this might account for a faster decay.
Alternatively, Kohno et al. (\cite{scuba:kohno05}) found that the flux
density at 93 GHz was approximately constant from 12 to 17 days 
after the burst before a rapid drop took place between 17 and 18
days after the burst.
It is therefore possible that a similar rapid drop took place 
$\la 16$ days after the burst at 850~$\mu$m.

The SCUBA flux density at 850~$\mu$m before the break in the
light curve is consistent with being constant.
Sheth et al. (\cite{scuba:sheth03}) found a small ($<20$\%) drop in the 
flux density at 100 GHz starting April 2 followed by a rise to its 
mean level on April 5.
This is similar to our observation on April 4 being lower than on 
April 3 and 5.
On April 5, the flux density at 43.3 GHz was the highest recorded,
and the flux density at the four other longer wavelengths measured 
was also temporarily higher on that day 
(Berger et al. \cite{scuba:berg03b}).
Thus this variability may have been real.
If so, it complicates the determination of the actual time for the break.

The observations during each SCUBA run show no significant variability.
One possible explanation for the fluctuations in the optical light curve 
is that the central engine had multiple ejections, rather than a single 
episode (e.g. Granot et al. \cite{scuba:gnp03}).
This would lead to refreshed shocks when the faster moving ejecta 
caught up with slower moving material.
A prediction of this model is that there should be short-lived
reverse shocks when the ejecta collide producing significant
longer wavelength flares.
The lack of large-scale variability at 850~$\mu$m agrees with the
results from other observations (Sheth et al. \cite{scuba:sheth03}),
and argues against this model.
However, we caution that our data only sparsely sample the light curve.

\subsection{Evolution of the longer wavelength spectrum}

Due to the poor observing conditions, the source was not detected 
at 450~$\mu$m in any of the individual observations.
The source was also not detected at 450~$\mu$m when the results from
separate days are combined.
The weighted mean for the first three days was $-69 \pm 33$ mJy
at 450~$\mu$m.

\begin{figure}
\centering
\includegraphics[bb=60 150 590 680,width=8.5cm,clip]{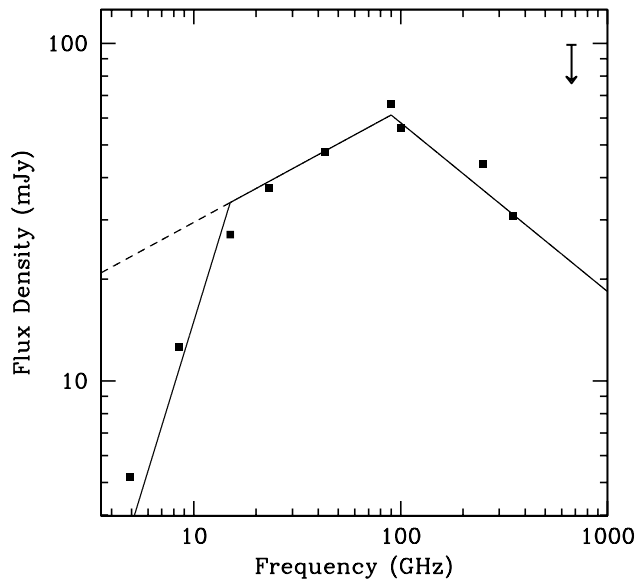}
\caption{The longer wavelength spectrum for GRB 030329 before the 
break in the light curve.
The data points show typical averaged values for $\sim 6$ days 
after the burst.
Added to the 850~$\mu$m (350 GHz) and 450~$\mu$m (670 GHz) data from
this paper are the results at
4.86, 8.46 and 15 GHz (Berger et al. \cite{scuba:berg03b}),
23 and 43 GHz 
(Berger et al. \cite{scuba:berg03b}; Kuno et al. \cite{scuba:kuno04}),
90 GHz (Kuno et al. \cite{scuba:kuno04}), 
100 and 250~GHz (Sheth et al. \cite{scuba:sheth03}).
The upper limit at 450~$\mu$m is $3 \sigma$.
The solid curve is $F_\nu~=~0.15~\nu^2$ up to 15 GHz,
$F_\nu~=~13.7~\nu^{1/3}$ up to a peak of 61 mJy at 90 GHz,
and $F_\nu~=~582~\nu^{-1/2}$ after the peak.}
\label{figure2}
\end{figure}

Figure~\ref{figure2} illustrates the longer wavelength spectrum before
the break in the light curve using the average fluxes for the observations 
$\sim 6$ days after the burst.
Since data has been combined from different observatories, and the source 
may be variable, the points will have uncertainties of up to $\sim 20$\%.
However, this does give a good qualitative picture for the longer 
wavelength spectrum at the time before the break.

The solid curve in Figure~\ref{figure2} shows one qualitative fit 
to the data.
The rise uses $F_\nu \propto \nu^2$ up to 15 GHz and
$F_\nu \propto \nu^{1/3}$ up to a peak of 61 mJy at 90 GHz.
The fall is $F_\nu \propto \nu^{-1/2}$.
The short dashed curve extrapolates the $\nu^{1/3}$ curve to highlight 
the break at longer wavelengths.

Spectral indices such as these are commonly found in synchrotron 
fireball models (e.g. Sari et al. \cite{scuba:spn98}; 
Piran \cite{scuba:piran99}; Wijers \& Galama \cite{scuba:wg99}; 
Chevalier \& Li \cite{scuba:cl00}; Granot et al. \cite{scuba:gps00};
Granot \& Sari \cite{scuba:gs02}; Panaitescu \& Kumar \cite{scuba:pk04}).
For example, the breaks could be $\nu_a = 15$ GHz and $\nu_m = 90$ GHz.

For the $R$ band ($4.2 \times 10^5$ GHz) this fit would extrapolate to 
give 0.9 mJy or $R \sim 16$.
The source was that bright $\sim 1$ day after the burst, and was
at $R \sim 18$ at 6 days after the burst.
In this case, it would then be necessary to have another break in the 
spectrum between the sub-millimeter and optical bands.
For example, this could be the cooling break $\nu_c$.

\begin{figure}
\centering
\includegraphics[bb=60 150 590 680,width=8.5cm,clip]{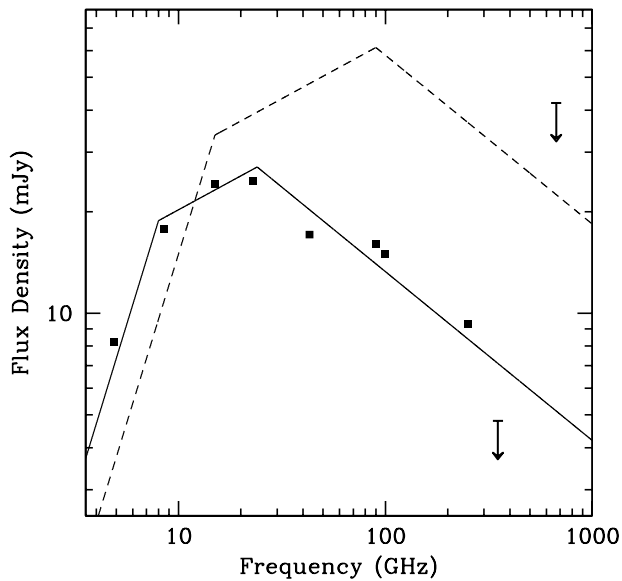}
\caption{The longer wavelength spectrum for GRB 030329 after the 
break in the light curve.
The data points show typical averaged values for $\sim 17$ days 
after the burst.
Added to the $3 \sigma$ 850~$\mu$m (350 GHz) and 450~$\mu$m (670 GHz) 
limits from this paper are the results at
4.86, 8.46, 15, and 43 GHz (Berger et al. \cite{scuba:berg03b}),
23 GHz (Kuno et al. \cite{scuba:kuno04}),
93~GHz (Kohno et al. \cite{scuba:kohno05}), 
100 and 250 GHz (Sheth et al. \cite{scuba:sheth03}).
The solid curve is $F_\nu~=~0.294~\nu^2$ up to 8 GHz,
$F_\nu~=~9.41~\nu^{1/3}$ up to a peak of 27 mJy at 24 GHz,
and $F_\nu~=~133~\nu^{-1/2}$ after the peak.
The short dashed curve shows the fit from Figure~\ref{figure2}.}
\label{figure3}
\end{figure}

Figure~\ref{figure3} illustrates the longer wavelength spectrum after
the break in the light curve using the average fluxes for the observations 
$\sim 17$ days after the burst.
Again, the uncertainties in the points may be up to $\sim 20$\%.
The solid curve in Figure~\ref{figure3} shows one qualitative fit 
to the data, with $F_\nu \propto \nu^2$ up to 8 GHz,
$F_\nu \propto \nu^{1/3}$ up to a peak of 27 mJy at 24 GHz, 
and a fall with $F_\nu \propto \nu^{-1/2}$.

The short dashed curve in Figure~\ref{figure3} shows the fit from 
Figure~\ref{figure2}.
Qualitatively, the spectral indices have stayed the same, with 
the two longer wavelength breaks ($\nu_a$ and $\nu_m$ in the
example above) moving to longer wavelengths and the flux at the 
peak falling.

However, the model predicts a flux density of 7.1 mJy at 850~$\mu$m
at $\sim 17$ days after the burst.
This is inconsistent with our lack of detection at the $3.7 \sigma$
level if we use the limit from April 15, and at the $4.4 \sigma$
level if we combine our last two SCUBA observations.

One way to reduce the disagreement between the model and the SCUBA
observation at $\sim 17$ days after the burst is if the third break 
($\nu_c$ in the example above) passed through the sub-millimeter band 
shortly before our last observations.
This could also help to explain why the decay of the 
850~$\mu$m light curve was faster than $\delta = 2$.
However, the spectral break would need to be sharp to fully explain
the SCUBA limit.
Also, in the simple fireball models, $\nu_c$ should remain constant after 
the jet break.
Thus this would require a more complex model, such as having the magnetic 
energy change with time (e.g. Yost et al. \cite{scuba:yost03}).

An alternative explanation for the lack of detection at 850~$\mu$m 
at $\sim 17$ days after the burst is that there could have 
been a rapid drop in the light curve, as discussed in \S 4.1.
The drop must have taken place earlier than 16.8 days after the burst
at 850~$\mu$m.
Figure~\ref{figure3} shows that the 250 GHz point was still on the 
$\nu^{-1/2}$ curve at 16.3 days after the burst.
The drop took place between 17 and 18 days after the burst at 93 GHz.
This indicates that if the drop is real, it did not occur at exactly
the same time for all wavelengths. 
Instead, it swept from shorter to longer wavelengths.

\subsection{SCUBA host galaxy limits}

The results in Table~\ref{table:host} show the weighted means for the 
last two days of SCUBA observations.
Although a small amount of afterglow emission may be present,
this contamination should be small.
There is no indication for any sub-millimeter emission from the host galaxy.

\begin{table}
\caption{Host galaxy fluxes for GRB 030329 determined by combining the 
SCUBA 850~$\mu$m and 450~$\mu$m observations taken on the last two days.}
\label{table:host}
\centering
\begin{tabular}{l c l l}
\hline\hline
Burst       & Redshift     & \multicolumn{2}{c}{Host flux density (mJy)} \\
            &              & 850 $\mu$m        & 450 $\mu$m      \\
\hline
GRB 030329  & 0.1685 & $-0.8 \pm 1.6$ & $17 \pm 14$ \\
\hline
\end{tabular}
\end{table}

It is not surprising that GRB 030329 is not a bright quiescent 
sub-millimeter source given that 
(1) it is at a low redshift, 
(2) the host is probably a dwarf galaxy, and 
(3) the X-ray and optical observations show relatively little excess 
absorption from dust above the Galactic value 
(Tiengo et al. \cite{scuba:tiengo03}; Matheson et al. \cite{scuba:mat03}).
However, given the short observing times and less than ideal observing 
conditions, the limits on the host galaxy fluxes are less restrictive than 
for some of the other bursts whose host galaxies have been observed by SCUBA
(Barnard et al. \cite{scuba:barn03}; Berger et al. \cite{scuba:berg03a}; 
Tanvir et al. \cite{scuba:tan04}; Smith et al. \cite{scuba:smith05}).

The SCUBA host galaxy compilation of Tanvir et al. (\cite{scuba:tan04})
focused on hosts with 850 $\mu$m rms flux densities $< 1.4$~mJy.
Since our rms is slightly higher than this, and the redshift of GRB 030329 
is $< 2$, the lack of detection of a sub-millimeter host galaxy does 
not shed any new light on the question of whether GRBs are closely 
linked to the most luminous dusty star-forming galaxies.

\section{Summary}

GRB 030329 was by far the brightest sub-millimeter afterglow seen to date.
Despite our limited observations, we were still able 
to constrain the break and/or decay light curve.
Assuming that the decay is not steeper than $\delta = 2$, 
the break must have taken place $\sim 7$ days after the burst
at 850~$\mu$m.
This jet break could have been at a slightly later time if another
break (perhaps the cooling break) passed through the sub-millimeter
band between 7 and 17 days after the burst, or if there was a drop
in the 850~$\mu$m flux at $\la 16$ days after the burst.

No short-lived large-scale brightenings were detected in the sub-millimeter 
light curve.
The flux density at 850~$\mu$m was consistent with being constant 
up to the jet break.
However, the 850~$\mu$m results also agree with those at longer 
wavelengths that show a slightly brighter flux $\sim 7$ days after 
the burst, right at the time of the break.

The peak of the afterglow emission was at $\sim 90$ GHz in the
days before the break in the light curve.
The spectral indices were qualitatively given by $+2$ and $+1/3$ 
up to the peak and $-1/2$ beyond it.
A simple modeling is consistent with the spectral indices remaining the
same as the afterglow evolved, with the spectral breaks moving to 
longer wavelengths at later times and the flux at the peak falling.

GRB 030329 did not have any significant sub-millimeter contribution 
from the host galaxy.

The observations of GRB 030329 show that for bright afterglows
it is important to monitor the flux evolution carefully to look
for fluctuations in the light curve and to determine the breaks.
Observations of new bursts are continuing to produce surprises, and 
there is much left to learn about GRB afterglows and host galaxies. 
To obtain a complete picture of their nature will require the careful 
study of many bursts to expand our sample.
Sub-millimeter observations with a $\sim$ mJy sensitivity are a key 
component to the multi-wavelength coverage.
To this end, our program of ToO observations using SCUBA is ongoing.

\begin{acknowledgements}

The James Clerk Maxwell Telescope is operated by The Joint Astronomy 
Centre on behalf of the Particle Physics and Astronomy Research Council 
of the United Kingdom, the Netherlands Organisation for Scientific Research, 
and the National Research Council of Canada.

We thank the JCMT Director Gary Davis for authorizing 
the ToO observations.
We are indebted to all the observers whose time was displaced by 
these observations, and acknowledge the dedicated efforts of the JCMT 
telescope operators Jim Hoge and Jonathan Kemp
for their valuable assistance with the observations.

We are grateful to Scott Barthelmy and Paul Butterworth for 
maintaining the GRB Coordinates Network (GCN), and to the other 
ground-based observers for the rapid dissemination of their 
burst results.

We acknowledge the data analysis facilities provided by the Starlink 
Project which is run by CCLRC on behalf of PPARC.

The work at Rice University was supported in part by 
AFOSR/NSF grant number NSF AST-0123487.

\end{acknowledgements}

\end{document}